# Blockchain-Powered Asset Tokenization Platform

Aaryan Sinha[1a], Raja Muthalagu[2b], Pranav Pawar[3], Alavikunhu Panthakkan[4c], and Shadi Atalla[5]

[1,2,3]Dept. of Computer Science, Birla Institute of Technology and Science Pilani, Dubai Campus, United Arab Emirates [4,5]College of Engineering and IT, University of Dubai, Dubai, United Arab Emirates Corresponding Authors:
[a]f20210183@dubai.bits-pilani.ac.in, [b]raja.m@dubai.bits-pilani.ac.in, [c]apanthakkan@ud.ac.ae

*Abstract*—Blockchain Technology has revolutionized Finance and Technology with its secure, decentralized, and trust-less methodologies of data management. In a world where asset value fluctuations are unprecedented, it has become increasingly important to secure one's stake on their valuable assets and streamline the process of acquiring and transferring that stake over a trust-less environment. Tokenization proves to be unbeaten when it comes to giving the ownership of one's asset, an immutable, liquid, and irrefutable identity, as of the likes of cryptocurrency. It enables users to store and maintain records of their assets and even transfer fractions of these assets to other investors and stakeholders in the form of these tokens. However, like cryptocurrency, it too has witnessed attacks by malicious users that have compromised on their very foundation of security. These attacks have inflicted more damage since they represent real-world assets that have physical importance. This project aims to assist users to secure their valuable assets by providing a highly secure user-friendly platform to manage, create and deploy asset-tokens, and facilitate open and transparent communication between stakeholders, thereby upholding the decentralized nature of blockchain and offering the financial freedom of asset ownership, with an added market value of a cryptocurrency-backed tokens.

## I. INTRODUCTION

### A. Tokens on a Blockchain

These are unique containers that are built on pre-existing blockchain networks. They were traditionally created for investors to back up projects as a sign of financial support. Their usage ever since has only expanded to make transacting on the blockchain much easier. They are easy to create and maintain, and help perform protocol-specific functions on Decentralized Applications. Their most popular use currently is to represent real-world assets (RWA) on the blockchain. Some examples of tokens include UNI, Binance, and popular NFT's like CryptoPunk and Bored Ape Yacht Club (BAYC) [1][3].

### B. Synthetic Assets

Assets can be classified as anything that adds monetary value to the owner in the long term. These asset classes popularly include real estate, jewelry, insurance, cash and even digital assets like data and cryptocurrency. Through the process of tokenization, we convert these assets to Synthetic Assets which are maintained on the token's blockchain and attain properties similar to cryptocurrency, like immutability and non-repudiation characteristics [2].

### C. Tokenization

It is the process of representing real-world or synthetic assets in the form of unique tokens. Fig. 1 lists the basic steps of tokenization. This simplifies the process of acquiring asset ownership, and even transferring that ownership to other people. Since it is set up on a blockchain, there is no need for a mediating body like a bank to carry out the transfer, thus increasing the security and trustworthiness of the ownership. It also beats the traditional approach by making typically hard-to-transact assets, like artifacts, easily liquid. Tokenization has seen its use in areas of all kinds of physical, digital, or financial assets, stablecoins and even in-game money [3].

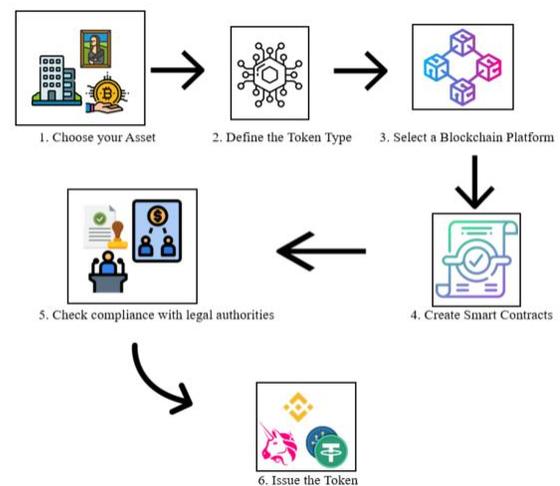

Fig. 1: Process of blockchain powered Tokenization

### D. The Problem to Solve

Assets currently face unprecedented changes in their valu-ation due to the high volatility of investment markets. Tradi-tional methods are also accompanied with a tedious process of acquiring and transferring these assets over the market to other stakeholders and investors, thus inhibiting investor participation, reducing security of acquiring ownership and subsequently decreasing asset value.

### E. Motivation

The market currently hosts an array of tokenization plat-forms for users to choose from. However, most of them have complex user interfaces and provide their resources at high initial prices with even more recurrent costs to achieve tokenization services. Since most platforms are organization centric, their control creates a level of centrality to the process and reduces the trust of the user on these services. These rea-sons suffice for the lack of participation and underperformance of these platforms.

F. Contributions

This paper attempts to combat the issues highlighted above and present a solution that enables efficient tokenization of these assets. The key contributions of this research are:

- Developing an open-source simulation-platform that al-lows common users to tokenize assets.
- Simplifying the process of liquidating assets and reducing any barriers.
- Making legal compliance more decentralized and efficient for users.

The remainder sections of this research paper entail the following: Section II elucidates the proposed solution and discusses its detailed workflow, and entails the results obtained of the demonstration. Section III elaborates on the literature review of related works, presents a comparison chart with ex-isting frameworks and platforms, and culminates this research paper with a note on the future scope of the solution.

## II. THE PROPOSED SOLUTION

This section discusses the development and implementation of the novel proposed solution, named WDApp, the Tokeniza-tion Distributed Application.

The solution presented in this paper attempts to develop a platform that is easily accessible to the public, providing them with flexible options to tokenize their assets. It ensures full security and transparency by using trusted blockchains like Ethereum. The platform will make use of Solidity based Smart Contracts to automate payments to investors and stake-holders. These smart contracts additionally can be embedded with legal token norms proposed by various institutions such as Securities and Exchange Commission (SEC) in order to ensure decentralized legal compliance and authentic token generation from the app. Decentralized Autonomous Organizations (DAOs) can further incorporate this solution with members updating contracts with local legal token norms and assisting users for a streamlined tokenization procedure.

A. Smart Contract Deployment

Tokenization is successfully able to automate token transfers and secure assets in a decentralized manner with the help of effective smart contracts. For the creation and representation of tokenized assets, the solution makes use of two tokenization contract standards, the ERC-20 for fungible tokens, and the ERC-721 for non-fungible tokens.

   a) The ERC-20 and ERC-721 Tokenization Standard: Tokens offer a lot of customizability and flexibility, allowing users to harness a variety of features to represent assets and transact on the blockchain. However, if these are too variant and have no common framework, it poses an issue with their inter-operability. This limits their compatibility and ease of transaction between contracts on different frameworks.
To address this, the proposed solution employs the Ethereum Request for Comment (ERC) tokenization standard for
building user-created tokens on the Ethereum chain. The solution constructs its contracts by harnessing the base structure of the ERC-20 and ERC-721 contracts in OpenZeppelin's Library.

   b) Deployment Procedure: As we demonstrate in Fig.2, the proposed contract workflow helps back any tokens minted to be secure and interoperable with other wallets and token frameworks. In addition, this contract is further customized to be able to mint more, burn and change ownership of these tokens, effectively making them highly scalable and integrable to any Ethereum-based Distributed Application (DApp).

B. The Web3 Script

The next phase of the solution is to incorporate the deployed contract in the back end of the DApp. The temporal relations shown in Fig. 2 explain how we obtain the contract binaries post deployment. Fig. 3 depicts state diagrams with the backend, Infura Application Program Interface (API), and the Streamlit-JavaScript front end integration, hence illustrating the complete full-stack working of WDApp.

C. Tokenization Platform Features

This section demonstrates the salient features of the WDApp tokenization platform through a series of images from Fig. 4 to Fig. 11. To effectively discuss the user-friendliness of the platform, we have depicted the user-journey through the ProjCoin token deployment process along with its symbol PBJ, and a sufficient arbitrary total supply of these tokens (Fig. 8). With its seamless integration with the back-end of the application (built on Web3.py), the user simply has to fill in a few details and can proceed to effortlessly transfer tokens (Fig. 4), check their balance (Fig. 5), let others spend tokens on their behalf (Fig. 6), and safely authenticate their transactions (Fig. 7). To explain the highly transparent and secure aspects of the platform, Fig. 9 to Fig. 11 show results of the validity of the token transfers along with the successful block generation of the transaction.

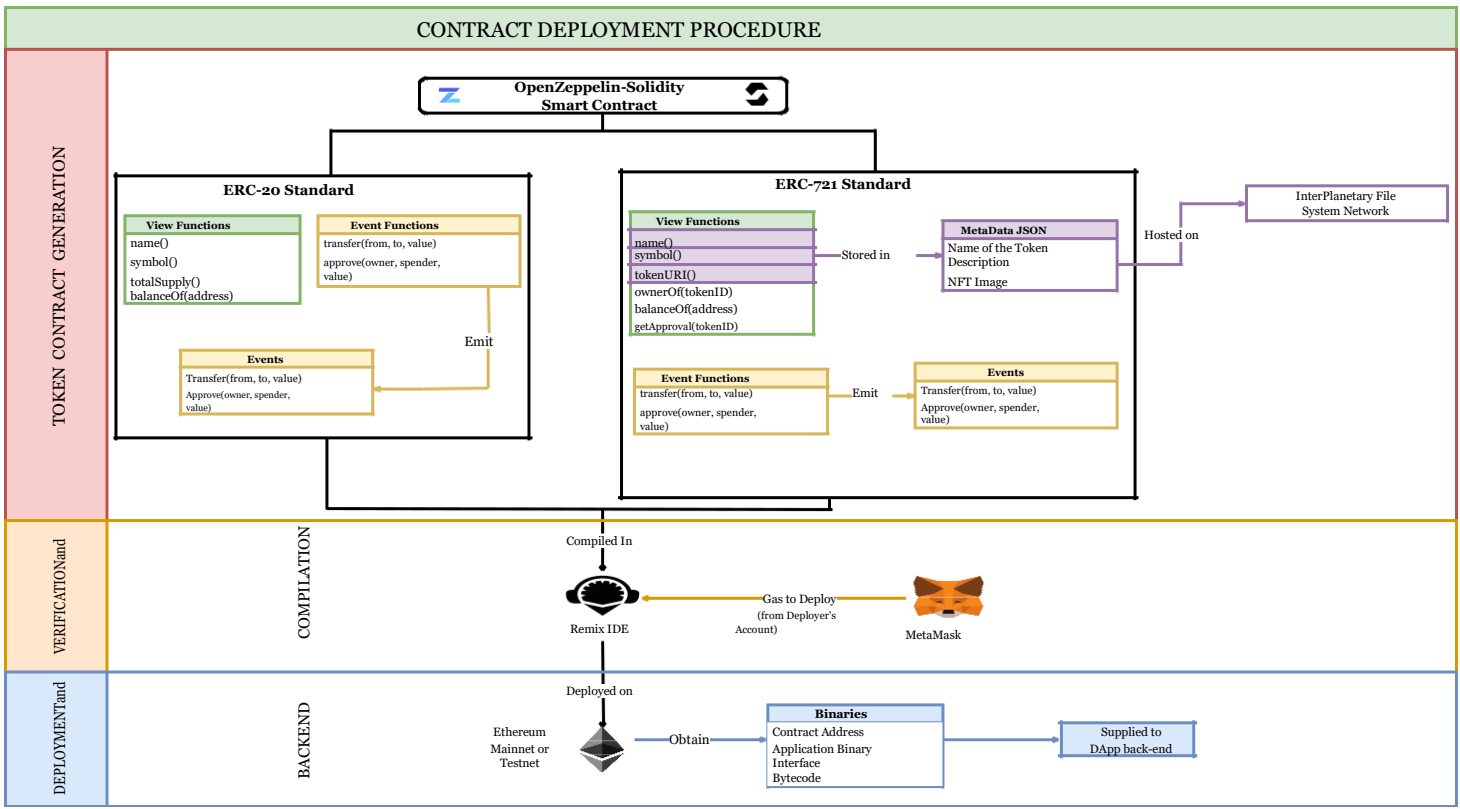

Fig. 2: A Swimlane and Class diagram depicting the Contract Deployment Workflow

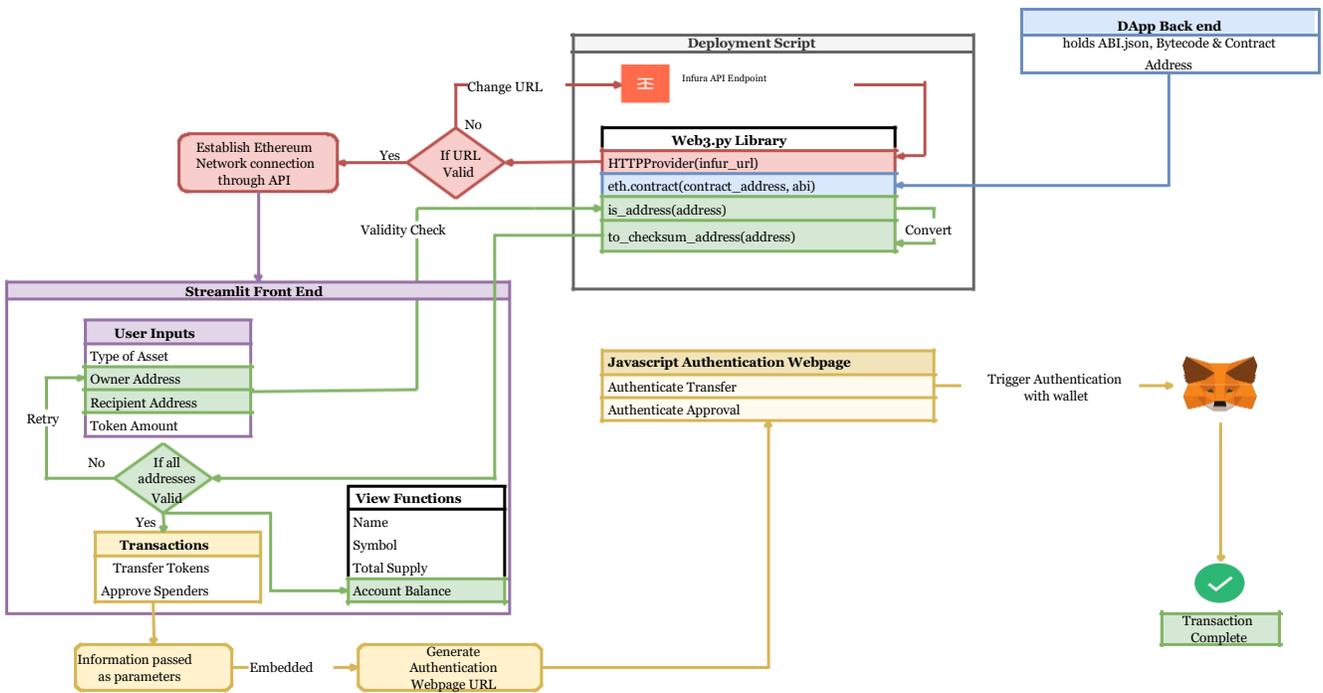

Fig. 3: State Diagrams illustrating the full-stack aspects of the distributed application workflow and end-user interaction with WDApp.

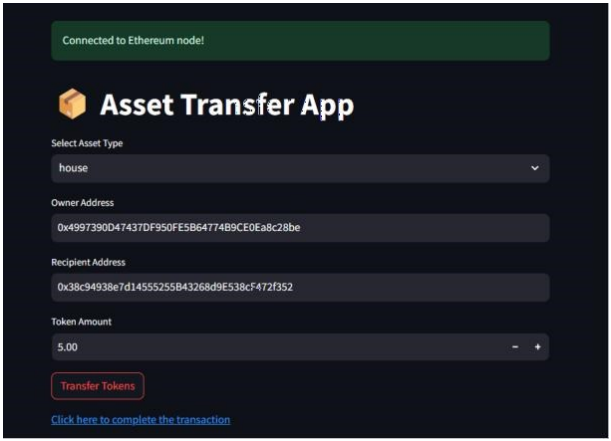

Fig. 4: Landing page with endpoint connected with the Transfer Token fields and functionality initiated.

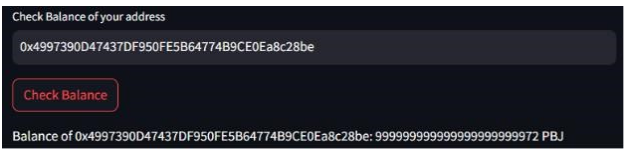

Fig. 5: Checking the balance number of tokens in an address after transfers.

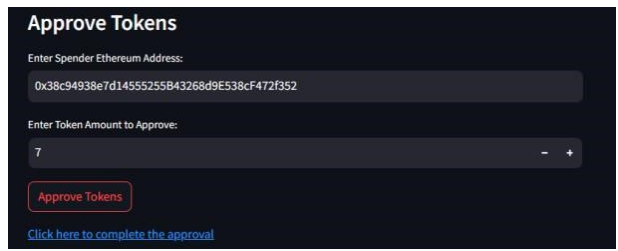

Fig. 6: Approve Tokens fields with functionality initiated.

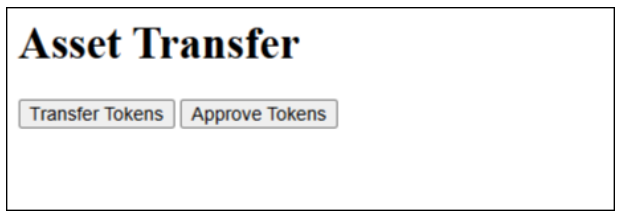

Fig. 7: Authentication JavaScript Webpage.

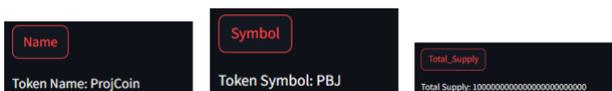

(a) Token name    (b) Token Symbol    (c) Total Supply

Fig. 8: View functions for token status.

The entirety of the back-end and the front-end code is open-sourced and can be found on the author's GitHub repository.

### D. Simulation environment

The proposed smart contract was deployed using the Remix Web IDE on a Sepolia Testnet. The proposed application was built on VS-Code version 1.89.1 and simulated on the Edge browser. The simulation was run on an AMD Ryzen 7 4700U system with a 64-bit Windows 11 operating system, 16GB RAM and a Radeon Graphics card.

### E. Results

The application's authenticity can be easily verified with the successful triggering of the MetaMask wallet (Fig. 9) upon initiating the transaction (Fig. 7). Upon completion, a transaction successful message pops-up in the browser, generating a unique transaction hash (Fig. 10) which can be pasted into any block-explorer, like Etherscan.io. Fig. 11 depicts the verified mining details on Etherscan, thus proving the validity of the generated nonce hash value. The speed of the process reflects the application's ability to safeguard user transactions, and the user's provision of sufficient gas to execute transactions.

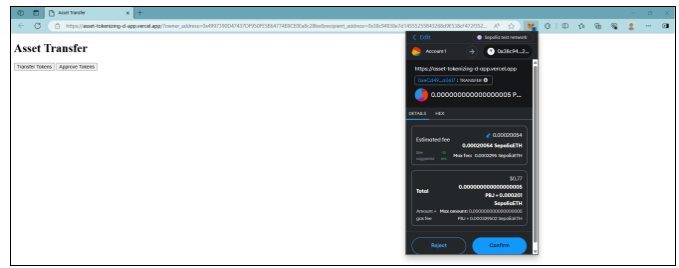

Fig. 9: Authentication Webpage trig-gers the MetaMask browser plug-in for wallet authorization.

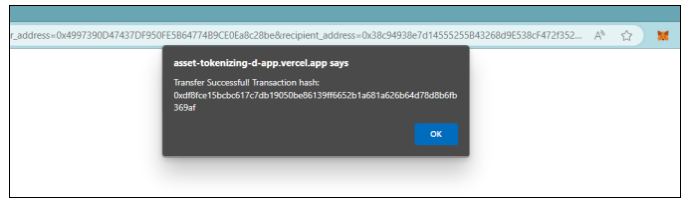

Fig. 10: Transaction Confirmation with generated Transaction Hash.

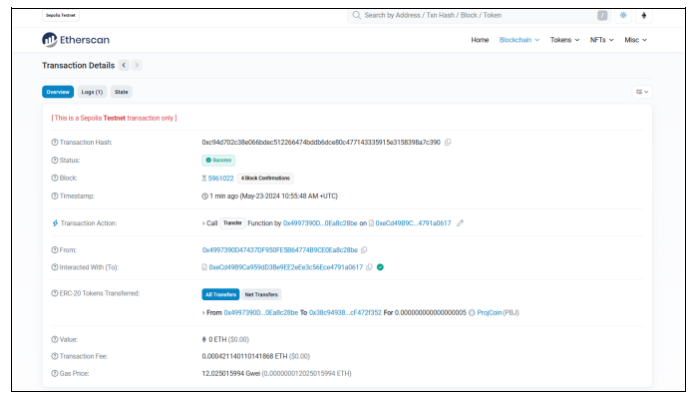

Fig. 11: Etherscan confirmation for successful mining of the transaction and addition to the blockchain.

## III. ANALYSES AND CONCLUSIONS

This section entails a thorough analysis conducted on vari-ous published papers on asset tokenization and presents their methodologies and drawbacks.

Frolov et al. [4] introduce a gold backed currency made by tokenization of real-world gold reserves to hasten and optimize the financial market turnover. However, the paper does not entail a user-friendly platform approach to facilitate the tokenization mechanism. Henker et al. [5] proposes a business model to manage the ownership and legality measures of real-estate in Germany by formulating a software-as-a-solution to tokenize assets and mapping legalities using digital-twin security tokens, but it does not explicitly mention a secure blockchain network to execute this. Shi et al. [6] foresee asset tokenization as the medium to democratizing investments to broaden and open access to the average individual. But the paper overlooks current De-Fi advancements and the use of Decentralized Autonomous Organizations (DAOs) to address legal issues. Zarifis et al. [7] present a thorough understanding of Non-Fungible Tokens (NFTs), their risks associated, marketplace understanding, and its use in incen-tivizing blockchain-based game development, but makes no mention of using the NFT business-model to tokenize real-world assets. Tian et al. [8] implement a novel tokenization approach to project finance and help bridge the gap between public and private sector's resources. The mechanism secures financial info, token-based investments and supports Public-Private Partnership (PPP), but employs SPVs to facilitate legal conformity of token finance with authorities, which adds centrality to the process. Buldas et al. [9] provide a pioneering research in tokenization by introducing unlimited framework scalability using Alphabill technology, along with their highly robust KSI Cash currency with the Alphabill Platform to acheive hyper-fast transaction throughput and a variety of features for cross-token transfers. The drawback here lies in the use of a less popular network which is yet to be evaluated for its ledger security and trust. Finally, Drogovoz et al. [10] present two case studies: Norilsk Nickel's Atomyze platform and The Circulor's tracking project in Rwanda, both of which paved the way for successful tokenization mechanisms for securing asset prices and tracking purchases and movemnts of materials. But the paper does not address scalability factors for real-world transaction volumes. Hence, it was evident from the literature review that the proposed solutions have either addressed the accessibility aspect of tokenization but not the scalability aspect, or vice versa, and seem to lack simplified user accessibility and enhanced security in their token platforms.

The proposed solution was then compared with existing plat-forms and contract frameworks to better assess its capabilities and address the low efficacy and adoption of these approaches. Table I presents a Comparative Chart on the qualitative analysis between other popular contract standards against the chosen standard, while Table II details the platform-specifc comparison between other competitor platforms and WDApp.

The analysis revealed that, due to the limited scalability of existing tokenizing frameworks and the complexity of such platforms, users were impeded from making blockchain-based investments, which explains the lack of its popularity. This paper proposes a solution to combat this problem by employ-ing a highly scalable token contract to efficiently secure and tokenize any real-world, financial, and digital asset, and also entails a novel and simplified user-application with minimum overheads and maximum decentralization for increased user participation in the blockchain asset market. The application however needs to be experimented with multiple real-time use cases and user-groups to ensure it achieves its goal of a transparent decentralized solution with rigorous testing across all Web 3.0 frameworks and asset markets.

TABLE I: Contract Standard, scalability and Token Compar-ison Table between various tokenization standards and the chosen standard (in pink and blue).

| Standard | Scalability | Sourcing | Fungible Assets | Unique Assets | Inter oper-ability |
|---|---|---|---|---|---|
| ERC-1155 [11] | Ethereum mainnet and testnet, only for tickets and game assets. | Open | Semi-Fungible, used for some trade-able game assets. | Unique game char-acter tokens. | Includes only few interop-erable functions. |
| BEP-20 [12] | Binance Smart Chain (BSC), for tradeable assets. | Closed | Only Binance sup-ported Assets. | No unique tokens | Only operable with the BSC frame-work. |
| KIP-37 [13] | For tradeable assets on the Klaytn Blockchain. | Open | Used to rep-resent utility tokens and stable coins. | No unique tokens | Only operable with Klaytn frame-work. |
| TRC-20 [14] | For tradeable assets on the TRON Chain. | Closed | Only TRON sup-ported Assets | No unique tokens | Only operable with TRON frame-work. |
| ERC-721 [15] | Most dominant NFT standard on Ethereum. | Open | Not Fungi-ble. | Can repre-sent any asset as an NFT. | Not Inter-operable but is used across the entire Ethereum frame-work. |
| ERC-20 [16] | Most widely used tradeable standard on Ethereum. | Open | Highest Fungi-bility for any real-world, digital, and financial asset. | No unique tokens | Operable with almost any Fungible standard. |

TABLE II: Platform Comparison Table analyzing the proposed solution (in blue) against Existing Tokenization Platforms to assess user convenience in the tokenization process.

| Platform | Onboarding | Overheads | Tokens Offered | Centrality |
|---|---|---|---|---|
| Fireblocks [17] | Only for enterprises and start-ups. | Higher prices for access to only a few supported tokens and chains. | No customizability of tokens. Only supports digital assets. | Token legality in hands of the platform, shifting control to legal parties. |
| Obito [18] | Easy to use Web UI. | Restricted to assets with Security Token Offerings (STO) on Swiss Bank ecosystem only. | Provides NFTs, and Security Tokens. | Bank ecosystem required for network deployment, shifting control to banks. |
| CodeFi [19] | Industry grade platform. Not user-friendly | Enterprise level initial investments needed. | No NFT support. | Governed by Consensys, users must adhere to compliance checks, shifts control of tokens to organization. |
| WDApp Platform | No onboarding needed. Plug-and-Play application, with MetaMask user authentication. | Minimum possible overheads. Only requires users to have enough tokens in wallet to transfer. | Represent any asset with Fungible and Non-Fungible Tokens. | Zero Centrality. Back-end is run on public and open-source blockchains (Ethereum). |

In the future, we aim to help users create Stable Coins and Security Tokens to better safeguard their tokenized assets from unprecedented market fluctuations. The open-source nature of the solution allows developers and DAO's to actively improve decentralized legal token compliance by incorporating Oracles and embedding Smart Contracts with essential norms such as Anti-Money Laundering (AML) and Know-Your-Customer (KYC) like uPort and SelfKey, and Zero-Knowledge Proofs (ZKP) that uphold user asset privacy for a hassle-free token generation.

ACKNOWLEDGEMENT

I thank the Director of BITS Pilani, Dubai Campus, for giving me a platform to develop my skills. I am grateful to Dr. Raja, Dr. Pranav and Dr. Alavikunhu , for the opportunity to create this project. I dedicate and owe this project to my family and friends.